\documentclass[11pt]{article}

\usepackage{amsmath,amssymb}
\usepackage{epsfig,wrapfig}
\usepackage[nospace]{cite}

\newcommand{\de}{\mathrm{d}}
\newcommand{\ee}{\mathrm{e}}
\newcommand{\im}{\mathrm{i}}
\newcommand{\ft}[2]{{\textstyle\frac{#1}{#2}}}

\begin{document}

\begin{titlepage}
\begin{center}
\hfill AEI-2005-173\\
\vskip 5em


{\Large \textbf{Partition function of 
 dyonic black holes in $\mathbf{N=4}$
    string theory}}
\vskip 15mm
\noindent
\textbf{J\"urg K\"appeli}\\[0.2cm]
{\em  
Max-Planck-Institut f\"ur Gravitationsphysik \\
Am M\"uhlenberg 1, 14476 Potsdam, Germany}\\[0.2cm]
{\tt kaeppeli@aei.mpg.de}
\end{center}
\vskip 4cm

\begin{center}
{\bf Abstract}
\end{center}
\begin{quotation}
  \noindent
  The dominant contribution to the semicanonical partition function of
  dyonic black holes of $N=4$ string theory is computed for generic
  charges, generalizing recent results of Shih and Yin. The result is
  compared to the black hole free energy obtained from the conjectured
  relation to topological strings. If certain perturbative corrections
  are included agreement is found to subleading order. These
  corrections modify the conjectured relation and implement covariance
  with respect to electric-magnetic duality transformations.
\end{quotation}

\vfill 

\noindent
{\textit{Contribution to the proceedings of the RTN
    conference on ``Constituents, Fundamental Forces and Symmetries of
    the Universe," September 2005, Corfu.}} 
\end{titlepage}

\section{Introduction}
\setcounter{equation}{0}

The conjecture of Ooguri, Strominger, and Vafa~\cite{Ooguri:2004zv}
relates the black hole partition function to that of topological strings.
In their proposal, the relevant black hole ensemble is the one in
which the magnetic black hole charges $p^I$ are treated
microcanonically while the electric charges $q_I$ are treated
canonically. This semicanonical partition sum is related to the
microcanonical partition sum $d(p,q)$ by a Laplace transform,
\begin{equation}
\label{eq:part}
  Z(p,\phi) = \sum_{q_I} d(p,q) \ee^{q_I \phi^I}\,,
\end{equation}
where the continuous variables $\phi^I$ are the electric potentials
conjugate to the quantized electric charges. When viewing $Z(p,\phi)$
as a holomorphic function in $\phi^I$, the black hole degeneracies
$d(p,q)$ can be retrieved by performing contour integrations as will
be reviewed. The conjecture amounts to comparing the black hole
partition function $Z(p,\phi)$ with the square of the topological
partition sum: $\ee^{\mathcal{F}(p,\phi)}=
\left|\ee^{F_{\mathrm{top}}}\right|^2$. Many encouraging results have
been presented to this extent~\cite{Dabholkar:2004yr,
  Aganagic:2004js,Verlinde:2004ck, Sen:2005pu, Dabholkar:2005dt}.  The
conjecture is, however, still lacking a precise formulation.
Modifications of the original conjecture are needed to implement
electric-magnetic duality covariance and duality symmetries such as S-
or T-duality. A comprehensive discussion will appear
in~\cite{cardoso:_black}.  Preliminary accounts of this work have been
presented at many occasions.\footnote{See, for instance, {\tt
    http://www.fields.utoronto.ca/audio/05-06/strings/wit/index.html}.}

In this paper, the partition function of $1/4$-BPS states is studied
that arises in $N=4$ compactifications of type-II string theory on
$K3\times T^2$. These models have a dual description in terms of
heterotic strings on $T^6$. For $1/4$-BPS states a formula for the
exact state degeneracy was proposed by Dijkgraaf, Verlinde, and
Verlinde~\cite{Dijkgraaf:1996it} and recently rederived by Shih,
Strominger, and Yin~\cite{Shih:2005uc}. It involves the automorphic
form $\Phi_{10}(\rho,\sigma,\upsilon)$ that transforms with weight 10
under the modular group $\mathrm{Sp}(2,\mathbb{Z})$. The arguments
$\rho$, $\sigma$, and $\upsilon$ form the period matrix of a Riemann
surface of genus~2. The state degeneracy of dyons depends only on the
$\mathrm{SO}(6,22)$-duality invariant products of the electric and
magnetic charge vectors,
\begin{equation}
\label{eq:charge-invariants}
    Q  = 2 q_0 p^1 + q^2\,,
    \qquad
    P  = - 2 q_1 p^0 + p^2\,,
    \qquad
    R  = q_0 p^0 - q_1 p^1 + p\cdot  q\,.
\end{equation}
Here, $p^2$ and $q^2$ are the contractions of $p^M$ and $q_M$ (with
$M=2,\ldots,27$) and involve a metric $C_{MN}$, which is related to
the intersection matrix on $K3$, and its inverse $C^{MN}$. Their
precise form will play no role. The charges can be identified with
those of D-branes wrapping the various cycles of $K3\times T^2$ and
with the quanta of winding and momentum of wrapped NS5-branes and
F-strings (see, for instance, \cite{Shih:2005uc} for details). The
dyon degeneracies $d(p,q)$ depend only on the invariants $Q$, $P$, and
$R$ and are given by the coefficients of the formal Fourier expansion of
$1/\Phi_{10}(\rho,\sigma,\upsilon)$. They can be extracted by
performing the contour integrals
\begin{equation}
\label{eq:dvvdegen}
  d(p,q) = \int \de\rho\,\de\sigma\de\upsilon
  \frac{
\ee^{\im\pi(Q\sigma+P\rho +(2\upsilon -1)
      R)}}
{\Phi_{10}(\rho,\sigma,\upsilon)}\,.
\end{equation}
In the limit of large charges, the logarithmic degeneracy agrees with
the entropy of the corresponding dyonic black holes, as was first
observed in~\cite{Dijkgraaf:1996it}. Subsequently, it was shown
in~\cite{Cardoso:2004xf} that the degeneracy formula precisely
captures both perturbative and non-perturbative corrections to the
Bekenstein-Hawking area law for black hole entropy, the origin of
which can be traced back to the presence of certain higher-derivative
curvature and non-holomorphic interaction terms in the effective
action. For the present set-up, the supergravity description was first
discussed in~\cite{LopesCardoso:1999ur}. As expected, the gravity
description reproduces only the semiclassical behavior
of~(\ref{eq:dvvdegen}). The analysis in~\cite{Cardoso:2004xf} shows
there are two type of corrections to the asymptotic density of states:
there are contributions that are exponentially suppressed in the limit
of large charges as well as perturbative corrections that are
subleading in this limit. In the following, the dominant contribution
to the black hole partition is evaluated. Here too, both type of
corrections will appear.

In~\cite{Shih:2005he}, Shih and Yin calculated the leading
contribution to $Z(p,\phi)$ for vanishing D6-brane charge $p^0$ and
determined the perturbative corrections in this limit. In this paper,
this computation is repeated for generic charges. While the presence
of a D6-brane charge does not lead to substantial technical
difficulties, it does uncover certain subtleties concerning subleading
terms of the measure. For large charges, the dominant contribution to
$Z(p,\phi)$ is of the form
\begin{equation}
\label{eq:partwithmeas}
  Z(p,\phi) \sim \sum_{k^I} \left[\sqrt{\Delta(p,\phi+2\pi\im k)}
    \ee^{\mathcal{F}(p,\phi+2\pi\im k)} + \ldots \right]\,.
\end{equation}
Here, $\mathcal{F}(p,\phi)$ is precisely the non-holomorphic
generalization of the free energy function given in
\cite{Cardoso:2004xf}. As discussed above, the ellipsis indicates that
microscopically one has exponentially suppressed corrections to the
leading contribution $\sqrt{\Delta} \ee^{\mathcal{F}}$.  These
originate from other rational quadratic divisors of
$\Phi_{10}$ and form the non-perturbative completion of the result.

The relevant contributions to the measure are accounted for by the
factor $\sqrt\Delta$. From electric-magnetic duality covariance one
expects that this factor is constructed from the determinant of a
generalized period matrix, and an argument is presented to this extent
in section~\ref{sec:macro}. The microscopic analysis shows that beyond
the subleading order the microscopic partition function
(\ref{eq:part}) differs from (\ref{eq:partwithmeas}). A extensive
discussion of these subtle issues will appear
in~\cite{cardoso:_black}.

Recently, Jatkar and Sen~\cite{Jatkar:2005bh} generalized the dyonic
degeneracy formula to a class of CHL-models and showed that,
asymptotically, it reproduces the entropy of the corresponding black
holes. The present set-up is a simple special case of these more
general models. The findings of this note can be generalized
to that class of CHL-models, as is discussed in~\cite{cardoso:_black}.

\section{Microscopic black hole partition function}
\setcounter{equation}{0}

In this section, the dominant contribution to the partition function
(\ref{eq:part}) is calculated, neglecting contributions that are
exponentially suppressed in the limit of large generic charges. Following
\cite{Shih:2005he}, the sum over $q_0$ and $q_1$ is converted into a
sum over invariants. From~(\ref{eq:charge-invariants}) it is clear
that only the combination $Q$ and $P$ can be used as independent summation
variables. The result is
\begin{equation}
  Z(p,\phi) = \frac{1}{p^1 p^0}\sum_{\phi^{0,1}\rightarrow\phi^{0,1}+
    2\pi i k^{0,1}}^{p^{1,0}-1}\sum_{q_M}\sum_{Q,P} d(p,q) \ee^{
    \frac{\phi^0}{2p^1}(Q-q^2) 
     - \frac{\phi^1}{2p^0}(P-p^2) + q\cdot\phi} \,,
\end{equation}
where $R$ is given by
\begin{equation}
R =  \frac{p^0}{2p^1}(Q-q^2)+ \frac{p^1}{2p^0}(P-p^2) + p\cdot q\,.
\end{equation}
There is a summation over imaginary shifts of $\phi^0$ and $\phi^1$
which is implemented by replacing $\phi^{0} \rightarrow\phi^{0}+ 2\pi
i k^{0}$ and $\phi^{1} \rightarrow\phi^{1}+ 2\pi i k^{1}$ in each
summand and, subsequently, by summing over the integers $k^{0,1} =
0,\ldots, p^{1,0}-1$. These shift sums enforce that only those
summands contribute for which $(Q-q^2)/2p^1$ and $(P-p^2)/2p^0$ are
integers.  Furthermore, they implement the required shift invariance
of $Z(p,\phi)$ under $\phi^{0,1}\rightarrow \phi^{0,1}+2\pi\im$. Using
the integral expression for the degeneracies~(\ref{eq:dvvdegen}), one
performs the sums over $Q$ and $P$. This yields the sums over
delta-functions $\sum_{n\in\mathbb{Z}}
\delta(\sigma-\sigma(\upsilon)+n)$ and $\sum_{m\in\mathbb{Z}}
\delta(\rho-\rho(\upsilon)+m)$, where
\begin{equation}
\label{eq:sigmarho}
  \begin{split}
   \sigma(\upsilon)&= -\frac{\phi^0}{2\pi\im p^1} - (2\upsilon-1)
   \frac{p^0}{2p^1}\,,\\
   \rho(\upsilon)&= \frac{\phi^1}{2\pi\im p^0} - (2\upsilon-1)
   \frac{p^1}{2p^0}\,.
  \end{split}
\end{equation}
These sums can be integrated against the contour integrals of $\sigma$
and $\rho$, which run in the strip $\sigma \sim \sigma +1$ and $\rho
\sim \rho +1$, with the result
\begin{equation}
    Z(p,\phi) = \frac{1}{p^1 p^0}\sum_{\phi^{0,1}\rightarrow\phi^{0,1}+
    2\pi i k^{0,1}}^{p^{1,0}-1}\sum_{q_M}
  \int\frac{\de\upsilon}{\Phi_{10}(\rho(\upsilon),\sigma(\upsilon),\upsilon)}
  \ee^{\im\pi\sigma(\upsilon) q^2+\im\pi\rho(\upsilon) p^2 +q_M \left( \phi^M +
      \im\pi(2\upsilon-1) p^M\right)} \,. 
\end{equation}
Note that in view of~(\ref{eq:sigmarho}) the integrand is invariant
under $\phi^{0,1} \rightarrow \phi^{0,1}+2\pi\im p^{1,0}$ as desired.
As pointed out by~\cite{Shih:2005he}, an extra phase factor $\exp[
-\im\pi R]$ is included in~(\ref{eq:dvvdegen}) relative to the
degeneracy formulae that appear
in~\cite{Dijkgraaf:1996it,Jatkar:2005bh}. In order to compare with the
macroscopic results it is useful to Poisson-resum with respect to
$q_M$. The result is
\begin{equation}
\label{eq:partpoisson}
   Z(p,\phi) = \frac{1}{p^1
     p^0}\sum_{\phi^{I}\rightarrow\phi^{I}+ 
    2\pi i
    k^I} \sqrt{\det{\im C_{MN}}}\int\frac{\de\upsilon\,\ee^{\frac{\im
        ( \phi+\im\pi(2\upsilon-1)p)^2 }{4\pi\sigma(\upsilon)}+
      \im\pi\rho(\upsilon) p^2}} {\sigma(\upsilon)^{(n-1)/2}
\Phi_{10}(\rho(\upsilon),\sigma(\upsilon),\upsilon)}\,,
\end{equation}
where $n=27$ for the present example, and the sum over $k^M$ is over
all integers. The shift-symmetry in $\phi^{0,1}$ is no longer obvious.

In a last step, the contour integral over $\upsilon$ is performed. The
contour runs horizontally in the strip defined by
$\upsilon\sim\upsilon+1$ and is confined to $\mathrm{Im}\rho\,
\mathrm{Im}\sigma > \mathrm{Im}\upsilon^2$, which for
(\ref{eq:sigmarho}) is given by $(a+b) \mathrm{Im} \upsilon< ab$ with
$2\pi a= \mathrm{Re}\phi^0/p^0$ and $2\pi b =- \mathrm{Re}\phi^1/p^1$.
One can show, using an $\mathrm{Sp}(2,\mathbb{Z})$ transformation,
that $\Phi_{10}(\rho,\sigma,\upsilon)$ is an even function in
$\upsilon$.  Using this and the periodicity $\upsilon \sim \upsilon +
1$, the $\upsilon$-contour can be closed thereby picking up the
encircled residues of $1/\Phi_{10}$. The result is twice the desired
integral.  In general, the discussion of the various contours in the
definition (\ref{eq:dvvdegen}) is subtle, since $\Phi_{10}$ has zeroes
even in the interior of the Siegel upper half plane. Fortunately, when
focusing on the leading contribution to $Z(p,\phi)$, these subtleties
do not play a role as long as the dominant residues are picked up. As
discussed in \cite{Dijkgraaf:1996it,Cardoso:2004xf,Shih:2005he}, the
leading contribution to the partition function comes from points that
lie on the rational quadratic divisor
\begin{equation}
  D = \rho\sigma -\upsilon^2 + \upsilon = 0\,.
\end{equation}
Around these points, $\Phi_{10}$ has the expansion (see \cite{Cardoso:2004xf}
for details)
\begin{equation}
\label{eq:degen}
  \Phi_{10}(\rho,\sigma,\upsilon)
 =\frac{
 \eta\left(\sigma'\right)^{24}
\eta\left(\gamma'\right)^{24}}{\sigma^{12}} \, {{D}^2}
+\mathcal{O}\left[D^4\right]\,, 
\end{equation}
where $\sigma'$ and $\gamma'$ are defined by
\begin{equation}
  \sigma' = -\frac{\rho}{\rho \sigma
    -\upsilon^2}\,,\quad \gamma' = \frac{\rho \sigma
    -\upsilon^2}{\sigma}\,.
\end{equation}
Inserting $\rho(\upsilon)$ and $\sigma(\upsilon)$ given in
(\ref{eq:sigmarho}) into these expressions one finds
\begin{equation}
  D  = (2\upsilon-1)
   \frac{\phi^0 p^1 - p^0\phi^1}{4\pi\im p^0p^1} + \frac{\phi^0
     \phi^1+\pi^2 p^1 p^0}{4\pi^2 p^1 p^0}\,. 
\end{equation}
The piece in $D$ quadratic in $\upsilon$ has canceled, and the
critical value $\upsilon_*$ is given by
\begin{equation}
(2\upsilon_* -1) = -\im \frac{\phi^0\phi^1+\pi^2 p^1p^0}{\pi(\phi^0p^1-\phi^1p^0)}\,.
\end{equation}
Therefore, the contour integral over $\upsilon$ is given by the
residue
\begin{equation}
\begin{split}
 Z(p,\phi) &=\sum_{\phi^{I}\rightarrow\phi^{I}+ 
    2\pi i 
    k^I} \sqrt{\det{\im C_{MN}}}\frac{(-8)\pi^3\im p^0
    p^1}{(\phi^0p^1-\phi^1p^0)^2}
\times\\
&\qquad \times
\frac{\de}{\de\upsilon}\left[  
  \frac{\sigma(\upsilon)^{12-(n-1)/2} \ee^{
\frac{\im}{4\pi\sigma(\upsilon)} \phi^2 + \im\pi
\left[\rho(\upsilon)-\frac{(2\upsilon-1)^2}{4\sigma(\upsilon)}\right]p^2
-\frac{2\upsilon-1}{2\sigma(\upsilon)} 
\phi\cdot
p}}{\eta(\sigma'(\upsilon))^{24}
\eta(\gamma'(\upsilon))^{24}}\right]_{\upsilon_*}+\ldots\,,
\end{split}
\end{equation}
where other exponentially suppressed contributions that come from
other divisors have been suppressed. The result takes the form
\begin{equation}
  \label{eq:partres}
  Z(p,\phi)= \sum_{ k^I} \mathcal{M}(p,\phi+2\pi\im k)
  \ee^{\mathcal{F}(p,\phi+2\pi\im k)} +\ldots\,.
\end{equation}
It is now shown that $\mathcal{F}(p,\phi)$ is exactly the
non-holomorphic generalization of the free energy given in
\cite{Cardoso:2004xf}. In addition, there is a measure factor
$\mathcal{M}(p,\phi)$, which is discussed below. To this extent the following
definitions are adopted:
\begin{equation}
\label{eq:defy}
  Y^I = \frac{\phi^I}{2\pi} + \frac{\im}{2} p^I\,,\quad \bar Y^I =
  \frac{\phi^I}{2\pi} - \frac{\im}{2} p^I\,, 
\end{equation}
which define the moduli $S = -\im Y^1/Y^0$, $\bar S = \im
\bar Y^1/\bar Y^0$, and $T^M = -\im Y^M/Y^0$, $\bar T^M= \im \bar
Y^M/\bar Y^0$. These relations are to be understood as defining
the quantities such as $S$ and $\bar S$ as functions of the complex
variables $\phi^I$. In particular, $S$ and $\bar S$, for instance,
are related by complex conjugation only if the
$\phi^I$ are real. In this sense one finds that on the divisor
\begin{equation}
4\rho_*-\frac{(2\upsilon_*-1)^2}{\sigma_*} =-
\frac{1}{\sigma_*} =  \im (S+\bar S)\,,
\quad
  \frac{2\upsilon_*-1}{2\sigma_*} = -\frac{\im}{2} (S-\bar S)\,,
\end{equation}
where $\rho_* = \rho(\upsilon_*)$ and $\sigma_*
=\sigma(\upsilon_*)$, and that $\gamma'(\upsilon_*) = \im S$ and $\sigma'(\upsilon_*)=\im
\bar S$.  
For $\mathcal{F}(p,\phi)$ in~(\ref{eq:partres}) these substitutions lead to
 \begin{equation}
\mathcal{F}(p,\phi)= (S+\bar S)\left[\frac{\phi^2}{4\pi} -\frac{\pi p^2}{4}\right] +
  \frac{\im}{2} (S-\bar 
  S)\phi\cdot p -\log\left[(S+\bar
    S)^{12}\eta(\im S)^{24}\eta(\im \bar{S})^{24}\right]\,.
\end{equation}
To arrive to this result, the factor $\sigma_*^{12}$ that arises
in~(\ref{eq:degen}) is absorbed into the exponent, while the factor
$\sigma_*^{-(n-1)/2}$ is a necessary part of the measure. The measure
factor $\mathcal{M}$ is given by
\begin{equation}
  \begin{split}
  \mathcal{M} = 4\pi^2 \sqrt{\det{C_{MN}}}(S+\bar S)^{(n-1)/2}
  &\left[-\frac{(T+\bar 
      T)^2}{2} + 2\pi (S+\bar 
  S)\left(12-\ft{n-1}{2}\right)\frac{(p^0)^2}{(\phi^0p^1-\phi^1p^0)^2}
  \right. \\ &\left.\qquad
    -\frac{12}{\pi(Y^0)^2}\partial_S\log \eta(\im S)
      -\frac{12}{\pi(\bar Y^0)^2}\partial_{\bar S}\log
  \eta(\im \bar S)   \right] \,.
  \end{split}
\end{equation}
Using that
\begin{equation}
  2\pi (S+\bar 
  S)\frac{(p^0)^2}{(\phi^0p^1-\phi^1p^0)^2} = -\frac{1}{2\pi(S+\bar
    S)}\frac{(Y^0-\bar Y^0)^2}{|Y^0|^4}\,,
\end{equation}
the measure can be rewritten as
\begin{equation}
\label{eq:resmeas}
  \begin{split}
 \mathcal{M} =4\pi^2\sqrt{\det{C_{MN}}}(S+\bar S)^{(n-1)/2}&
 \left[-\frac{(T+\bar T)^2}{2} +\mathcal{D}\Omega
   +\frac{(n-1)}{4\pi(S+\bar S)}\frac{(Y^0-\bar Y^0)^2}{|Y^0|^4}
 \right.\\ 
&\left.\qquad + \frac{36}{2\pi}\frac{1}{(S+\bar S)|Y^0|^2} \right]\,,
  \end{split}
  \end{equation}
where the operator $\mathcal{D}$ is given by 
\begin{equation}
\begin{split}
\mathcal{D} = \frac{2}{(Y^0)^2}\partial_{S}+ \frac{2}{({\bar
      Y}^0)^2}\partial_{\bar{S}}-\frac{2(S+\bar
    S)}{|Y^0|^2}\partial_S\partial_{\bar S}
\end{split}
\end{equation}
and $\Omega$ is the same function that appeared in~\cite{Cardoso:2004xf}:
\begin{equation}
\Omega = -\frac{6}{\pi}\log\eta(\im
  S) -\frac{6}{\pi}\log\eta(\im
  \bar S)-\frac{3}{\pi} \log[S+\bar
    S]\,.
\end{equation}
This completes the computation of the dominant contribution to the
semicanonical black hole partition function. The expressions
$\exp\mathcal{F}(p,\phi)$ and $\mathcal{M}(p,\phi)$ coincide with the
expressions found in~\cite{Shih:2005he} in the limit $p^0\rightarrow
0$, while, not surprisingly, the ranges of the sums over $k^0$ and
$k^1$ are different.

In the next section it is argued that the first two terms in the
bracket of (\ref{eq:resmeas}) are to be treated as the leading terms
and that this gives rise to a precise agreement with the leading
perturbative corrections induced by the measure factor $\sqrt\Delta$
of (\ref{eq:partwithmeas}).  The third, $n$-dependent term in
(\ref{eq:resmeas}) is not T-duality invariant and its presence
reflects the fact that for the set-up discussed here (\ref{eq:part})
breaks T-duality invariance. Both this and the forth term in
(\ref{eq:resmeas}) are not captured directly by the approach discussed
in the following.

\section{Semiclassical black hole partition function and duality}
\label{sec:macro}
\setcounter{equation}{0} The Ooguri-Strominger-Vafa proposal
\cite{Ooguri:2004zv} must be modified in order to ensure the
covariance with respect to electric-magnetic duality transformations
and, in particular, to obtain $S$- and $T$-duality invariant
results~\cite{cardoso:_black}. These modifications should account for
the leading perturbative corrections calculated in the previous
section.  One way to derive these modifications is to start
from a symplectically covariant expression for the black hole
partition sum,
\begin{equation}
\label{eq:covpart}
Z(\chi,\phi) = \sum_{p,q} d(p,q) e^{q_I\phi^I-p^I\chi_I}\,.
\end{equation}
The additional sum over the magnetic charges is weighted by the
magnetic potentials $\chi_I$. The electric and magnetic potential
$(\phi^I,\chi_I)$ transform as a vector under electric-magnetic
duality transformations. Assuming that the microscopic degeneracies
transform as a function [as is expected to be the case for
(\ref{eq:dvvdegen})], the left-hand side of~(\ref{eq:covpart}) is
invariant under symplectic transformations. By definition, one has
$Z(\chi+2\pi\im,\phi) = Z(\chi,\phi+2\pi\im) =Z(\chi,\phi)$.  Viewing
$Z(\chi,\phi)$ as a holomorphic function in $\chi_I$ and $\phi^I$, the
degeneracies $d(p,q)$ or $Z(p,\phi)$ can be retrieved by an inverse Laplace
transform. For example,
\begin{equation}
\label{eq:degfrompart}
  d(p,q) =\prod_{I,J}\frac{1}{(2\pi\im)^2} \int \de \chi_I\,\de\phi^J
  Z(\chi,\phi) \ee^{-q_K\phi^K+ p^K\chi_K}\,,
\end{equation}
where contours run in the strips $\chi_I\sim\chi_I+2\pi\im$ and
$\phi^J\sim\phi^J+2\pi\im$. Of course, it would be desirable to derive
$Z(\chi,\phi)$ directly from a degeneracy formula such as
(\ref{eq:dvvdegen}), but this seems difficult. 

Inspired by~\cite{Ooguri:2004zv}, a symplectically covariant function
$Z(\chi,\phi)$ is suggested in~\cite{cardoso:_black} that reproduces,
using~(\ref{eq:degfrompart}), the expected black hole entropy in the
semiclassical regime of large charges. The existence of such a
function is intimately related to existence of a variational principle
for black hole attractors and black hole entropy. The leading
contribution to $Z(\chi,\phi)$ is of the form
\begin{equation}
\label{eq:parthesse}
  Z(\chi,\phi)\sim \sum_{l_I,k^J} \ee^{2\pi\mathcal{H}(\chi+2\pi\im
    l,\phi+2\pi\im 
    k)}+\ldots\,,
\end{equation}
where $\mathcal{H}(\chi,\phi)$ is a generalized version of the Hesse
potential and includes the effects of higher-derivative curvature
interactions and possibly of non-holomorphic corrections. The sums
over $l_I$ and $k^J$ are expected to be present and reflect the
fact that $\mathcal{H}(\chi,\phi)$ generically does not have any
periodicity properties in $\chi$ and $\phi$. The example of the
previous section shows that the ranges of some of the summations can
be restricted even though the corresponding periodicities might become
apparent only after resummation. The ellipsis indicates that, similar
to~(\ref{eq:partwithmeas}), one expects exponentially suppressed
contributions that form the full non-perturbative completion of the
expression. The assumption~(\ref{eq:parthesse}) is rather compelling:
inserting~(\ref{eq:parthesse}) into~(\ref{eq:degfrompart}) and
performing a saddle-point approximation with respect to both the
electric and magnetic potentials, one finds that this
semiclassical result is in precise agreement with the general black
hole entropy formula. Clearly, when comparing with microscopic entropy
formulae, corrections to this semiclassical black hole entropy
arise~\cite{Cardoso:2004xf,Jatkar:2005bh}. Such effects lead to
additional subleading contributions to~(\ref{eq:parthesse}) and are
discussed in~\cite{cardoso:_black}.

In order to make a connection with~(\ref{eq:partwithmeas}) one can now
use~(\ref{eq:parthesse}) as the starting point and perform an inverse
Laplace transform with respect to the magnetic potentials only. For
generic directions $\chi_I$, the sums over the shifts combine with the
integrals along the strips $\chi^I\sim \chi^I+2\pi\im$ to give
contours running parallel to the whole imaginary axis. When performing
these integrals in saddle-point approximation one recovers precisely
(\ref{eq:partwithmeas}), where $\Delta$ is given
by the determinant of the period matrix that is suitably generalized
to include certain higher-derivative curvature interactions and
non-holomorphic corrections. For the set-up discussed in the previous
section, the result is given, up to a numerical factor, by the sum of
two squares:
\begin{equation}
 \label{eq:delta}
  \Delta(p,\phi) \sim \det{C_{MN}} (S+\bar S)^{n-1}
  \left[\left(-\frac{1}{2}(T+\bar T)^2+\mathcal{D}\Omega\right)^2-
  \frac{4(S+\bar S)^2}{|Y^0|^4}|D_S\partial_S\Omega|^2\right]\,,
\end{equation}
where
\begin{equation}
  D_S\partial_S\Omega = \left(\partial_S^2+\frac{2}{S+\bar
    S}\partial_{S}\right)\Omega\,. 
\end{equation}
Up to an overall rescaling by $|Y^0|^4(S+\bar S)^2$, the two terms in
the bracket (\ref{eq:delta}) are each invariant under S- and T-duality
transformations. In order to relate this to the microscopic result, one
compares $\log\sqrt{\Delta}$ with $\log \mathcal{M}$ given by
(\ref{eq:resmeas}). Thereby, one treats the first term in the bracket
of (\ref{eq:delta}),
\begin{equation}
  (S+S) |Y^{0}|^2 \left(-\frac{1}{2}(T+\bar T)^2+\mathcal{D}\Omega\right)\,,
\end{equation}  
as the leading, duality invariant part and expands $\log\sqrt\Delta$
in inverse powers of this quantity. The same is done for the
expression $\log\mathcal{M}$ given in (\ref{eq:resmeas}) and one finds
precise agreement to leading order in these expressions. The two
partition functions therefore agree to subleading order. Beyond this
order there are, not unexpectedly, certain deviations. A discussion of
the origins of these effects is given
in~\cite{cardoso:_black}.

\vskip 1cm

\noindent
This work is partly supported by the EU contract MRTN-CT-2004-005104.

\providecommand{\href}[2]{#2}\begingroup\raggedright\endgroup


\end{document}